\documentclass[graybox, natbib, nosecnum, twocolumn]{svmult}
\bibpunct{(}{)}{;}{a}{}{,} 

\usepackage{mathptmx}       
\usepackage{helvet}         
\usepackage{courier}        
\usepackage{type1cm}        

\usepackage{makeidx}         
\usepackage{graphicx}        
\usepackage{multicol}        
\usepackage[bottom]{footmisc}
\usepackage[normalem]{ulem}	
\usepackage{hyperref}  
\usepackage{soul}   

\usepackage{multirow}
\usepackage{booktabs}
\usepackage{enumitem}
\usepackage{comment}

\begin{document}

\title*{Security Reputation Metrics}
\author{Maciej Korczy\'nski \thanks{Univ. Grenoble Alpes, CNRS, Grenoble
INP, LIG, France} and Arman Noroozian \thanks{Delft University of Technology, Netherlands}} 
\authorrunning{Korczy\'nski and Noroozian}
\institute{Maciej Korczy\'nski \at Univ. Grenoble Alpes, CNRS, Grenoble INP, LIG, France, \email{maciej.korczynski@univ-grenoble-alpes.fr}
\and Arman Noroozian \at Delft University of Technology, Netherlands \email{a.noroozian@tudelft.nl}}
\maketitle

\section{Definitions}
Security reputation metrics (\textit{aka.} security metrics) quantify the security levels of organization (e.g., hosting or Internet access providers) relative to comparable entities. 
They enable benchmarking and are essential tools for decision and policy-making in security, and may be used to govern and steer responsible parties towards investing in security when economic or other decision-making factors may drive them to do otherwise.

\section{Background}
We increasingly interact with online digital content, which 
relies on services provided by so-called \textit{Internet intermediaries}, among them Internet service providers
(ISPs), hosting providers, domain name registrars, search engines, payment providers, certification authorities, cloud service providers, social network operators and e-commerce suppliers. Simultaneously, miscreants have been abusing the products and infrastructure of such service providers towards cybercrime by compromising their security and using them in unintended ways. 

A wealth of research into cybercrime points to how cybercriminals misuse hosting services \citep{Nikiforakis2011,Noroozian2019}, domain names \citep{Felegyhazi2010a,Liu2011,Hao2011,Hao2013,Szurdi2014,PochathMGPDJK20}, DNS services \citep{Bilge2011,Canali2011} and mail servers \citep{Stone-Gross2011a,Levchenko2011} to name a few examples.

A typical situation with intermediary services is that their consumers are at an inherently disadvantaged knowledge position in which they don't know much about how secure these services are. In contrast the service providers themselves know much more. 
Without additional security-related information, users typically base their decision to subscribe to a particular service on more readily available information, for instance pricing in case of hosting, or available bandwidth in case of broadband ISP. 
In other words, it is difficult for other businesses, consumers and regulators to reliably assess how secure intermediary services are. 
This so-called \textit{information asymmetry} about the security of intermediary services, combined with the fact that typically, parties other than the intermediaries themselves bear the cost of the cybercrime enabled through their services \citep{Anderson2012}, leads to an erosion of their incentives to adequately invest in security. 
In other words their economic incentives are misaligned with security goals.

Thus, systematically comparing the security performance of digital services, and intermediaries, may help reduce the security information asymmetries from a consumer perspective. 
Security reputation metrics are essential to this end and may help in reducing cybercrime, which is as much a technical issue as a problem of economic incentives \citep{Anderson2001}. 
Various stakeholders such as network service providers \citep{Asghari2015b}, domain registries \citep{tld_maciej}, law enforcement agencies \citep{Noroozian2015}, and even policy makers \citep{sadag} employ security metrics to answer questions like which are the worst service providers and what actions should be taken to steer market driven economies towards improved security outcomes. 

Existing metrics typically compare security based on either (\textbf{i}) how frequently abuse incidents occur (or vulnerabilities are discovered), i.e. are based on counting the number of incidents (vulnerabilities), or (\textbf{ii}) based on how timely incidents are remediated once they have occurred. 
The number of maliciously registered domain names, compromised end-user machines, and machines running outdated software per service provider are examples of the former case. 
The amount of time required to remediate, block and remove phishing or malware spreading web-pages are examples of the second metric type. 

Count based metrics are typically normalized by estimates of the size of each intermediary's potential \textit{attack surface} to control for more exposed intermediaries that have higher probability of experiencing incidents. This enables apples-to-apples comparisons between intermediaries of various exposure. The number of advertised IP addresses by a hosting provider, or the number of domains that it hosts, for instance may be used to estimate a hosting provider's potential attack surface \citep{Noroozian2015,Tajalizadehkhoobnoms, Tajalizadehkhoobtoit}.

\section{Application}

The following subsections present examples of security reputation metrics for different types of providers and show how they are used by Internet stakeholders in reducing cybercrime and aligning economic incentives with better security.

\subsection{Hosting Providers}

Hosting providers are companies that provide servers via which customers can make content or services available on the Internet e.g. websites, email or even sharing of files. As with all services on the Internet, they are also abused for criminal purposes. 
Think of phishing sites, command-and-control servers for botnets, child sexual abuse material (CSAM), malware distribution, and spam servers.

In theory, hosting providers can mitigate or prevent the abuse of their infrastructure by following security best practices set forth by organizations like the Messaging, Malware and Mobile Anti-Abuse Working Group (M3AAWG). For example, by vetting customers, monitoring their infrastructure for signs of compromise or even running anti-virus software to name a few recommended practices. Yet following such advice remains voluntary, and there is considerable variation in how hosting providers choose to act when it comes to abuse and securing their infrastructure. 
Naturally then, some providers are abused more often than others.  

Therefore, the question of which hosting providers are better at securing their infrastructure, is one that may be answered through security metrics that quantify how effective each provider is at curbing abuse. 

A systematic approach to metrics development within the hosting market is presented in a study by \citeauthor{Noroozian2015}, which is scoped to the Netherlands and the result of a collaboration between several local authorities. The aim of the study being to answer the following question: Which are the worst hosting providers in Dutch jurisdiction? Metrics developed within this study were used to steer the Dutch hosting market towards more effective security practices via involving several local authorities \citep{Noroozian2015}.
Furthermore, \citeauthor{Noroozian2017}, have developed metrics for comparing the security of hosting providers globally \citep{Noroozian2017}, which are shown to have considerable predictive power. Other work discusses how multiple approaches to curbing abuse within the hosting market, from loose market-based approaches, to stricter regulation-based approaches, may benefit from the employment of security metrics \citep{Fryer2015, Noroozian2020a}.

\subsection{Top-Level Domains}

There exists little empirical information about the security of entire Top-Level Domains such as .com, .nl,  or .top.
\cite{tld_maciej} were first to present security metrics for this ecosystem and have measured their operational values.  
They compared entire TLDs against the rest of the market. However, they have explicitly distinguished
those metrics from the  
objective of measuring the security performance of the registry operators. The reason is that a TLD is not a single organization but
constitutes an entire `domain name ecosystem' of different
types of intermediaries such as domain registries, registrars, or hosting providers.

The follow-up study requested by ICANN, spanning the period up to the end of 2016, investigated the following research question: How do abuse rates in the new gTLDs (e.g., .top, .science, .bank, or .site) compare to legacy gTLDs (e.g., .com, .net, .edu, or .org), since the introduction of the new gTLD program in 2013?
To determine the distribution of abusive activities across the gTLDs, \citeauthor{Korczynski2018} have analyzed the number of reported domains from reputed URL and domain blacklists normalized by the size of their respective TLD, calculated as the number of 2\textsuperscript{nd}-level domain names present in a zone file for each gTLD. 

They have compared abuse rates separately for compromised and maliciously registered domains \citep{comar}.
Reputation metrics reflecting spam activity in the new and legacy gTLDs have revealed an interesting trend: miscreants seem to be switching from abusing legacy to new
gTLDs when it comes to maliciously registered spam domains. 
In the last quarter of 2016, new gTLDs collectively had approximately one order of magnitude higher rate of spam domains per 10,000
registrations compared to legacy gTLDs.
Moreover, as many as 15 most abused new gTLDs had more than 10\% of all registered domain names blacklisted by Spamhaus at the end of 2016.
Finally, as many as 51.5\%, 47.6\%, and 33.4\% of all .science, .stream, and .study new gTLDs, respectively, were maliciously registered by cybercriminals and blacklisted by Spamhaus.

ICANN has used the calculated reputation metrics to review the existing anti--abuse safeguards in new gTLDs, and to introduce more effective ones before an upcoming new gTLD rollout.

\subsection{Internet Service Providers}
A significant amount of scientific work addresses the role of Internet Service Providers (ISPs)---network access providers---in mitigating cybercrime, for instance, through botnet mitigation by employing security metrics \citep{Eeten2010,Asghari2015a,EetenLMAK16}. 
ISPs typically provide Internet connectivity to customers and thus are in a unique position to mitigate certain forms of cybercrime at their origin. 

ISPs should also follow security best-practices 
to mitigate abuse.
Examples of some security best-practices for ISPs include the use of walled-gardens to quarantine and isolate infected machines connected to the Internet \citep{CetinGATE18,CetinGAKITTYE19}, or deploying Source Address Validation, also known as BCP38, to prevent Distributed Denial-of-Service (DDoS) attacks from being launched via their infrastructure \citep{LuckieBKKKc19,closed,loops}.
Yet, again, the voluntary nature of implementing such best practices results in certain ISPs experiencing a higher level of abuse than others due to having laxer security practices. 

Differences among ISPs in mitigating botnet infections, for instance, have been quantified in several studies. 
\citeauthor{Eeten2010} , for example, found that just 50 ISPs account for over half of all spam sources suggesting concentrations of spambots within a few ISPs worldwide. 

A typical approach in such studies is to first process global or national datasets of botnet activity from available sinkholes 
and to extract IP addresses of infected end-user machines. 
The methods employed typically map each bot-infected IP address to an ISP and then counts the IP addresses seen in each ISP 
per day to account for IP churn \citep{MouraGLPAE15}.
Security metrics to compare botnet mitigation among ISPs are then calculated by dividing this count by the numbers of subscribers of each ISP, where larger number indicate less effective mitigation by the ISP. 
Such security metrics for botnet mitigation among ISPs have been employed successfully to incentivize ISPs, within the Netherlands for example, to subscribe to threat intelligence data feeds, and deal with bot infections within their networks \citep{EetenLMAK16}.

\section{Open Problems and Future Directions}
Empirical measurements and analysis of security indicators leading to reliable security reputation metrics has proven to be quite challenging. 
The challenge partly lies in limitations of data: e.g. coverage, measurement errors and biases that are invariably linked to metric limitations. 
For example, the construction of security metrics typically depends on 
chaotic internet operations data such WHOIS information that are error prone and incomplete, or on ever evolving dynamic BGP routing data for attributing security incidents to the responsible entities. 

The security incidents themselves are observed through various opaque abuse feeds with no clear documentation of their collection methodologies, accuracy, and biases. Abuse feeds contain various degrees of false positive incident information or carry biases that are not well documented or understood. 
Abuse feed coverage is also limited with an unknown number of security incidents goes unnoticed as false negatives for each entity thereby affecting metric outcomes. 

In addition to limitations in data, methodological challenges in constructing metrics also exist. 
A particularly challenging methodological aspect is that of identifying service providers within certain markets. For example, in the case of hosting providers, there is no maintained authoritative list of companies, even at a country level, to identify the companies that offer hosting services. This problem is worsened by layers of smaller companies that resell the services of larger hosting providers.
Similar problems exist for domain registrars. Yet, such information is vital for the construction and comparison of service providers against their competitors at a market level to make security metrics more useful. 

Another methodological challenge is an incomplete causal understanding of the factors that drive abuse across online services. A better and more complete causal understanding of such drivers enriches and allows for the construction of security metrics that are better interpretable, easier to understand, and more useful.

Finally, there are practical limitations as well including the fact that metrics do not reflect the intent of bad service providers whether it be negligent or criminal behavior for instance in the case of bullet-proof services that cater to cybercriminals with the promise of ignoring or delaying lawful take down requests.  
As such, security metrics can be gamed by coordinated criminals and thus there are limits to how they may be interpreted.

\bibliographystyle{spbasic}  
\bibliography{SRM.bib}

\end{document}